\documentclass[aps,prb,epsfig]{revtex4}

\begin{document}

\title{ Nonequilibrium Statistical Mechanics and Thermodynamics from Darwinian Dynamics:
 a Primer }

\author{ P. Ao }

\address{ Department of Mechanical Engineering,
          University of Washington, Seattle, WA 98195, USA }

\date{December 27 (2005)}

\begin{abstract}
We present here an exploration on on the physical implications of
the Darwinian dynamics. We first show that how the nonequilibrium
statistical mechanics emerges naturally. We then show that the
first three laws of the thermodynamics, the Zeroth Law, the First
Law and the Second Law can be followed from the Darwinian
dynamics, except the Third Law. The inability to derive the Third
Law indicates that the Darwinian dynamics belongs to the
"classical" domain. Specifically, the Second Law is proved from
the dynamical point of view. Two types of current dynamical
equalities are explicitly discussed in the paper: one is based on
Feynman-Kac formula and one is a generalization of the Einstein
relation. Both are directly accessible to experimental tests. Our
demonstration indicates that the Darwinian dynamics is logically a
simple and straightforward starting point to get into
thermodynamics and is complementary to the conservative dynamics
dominated in physics.
 \\
 PACS numbers: \\
05.70.Ln
 Nonequilibrium and irreversible thermodynamics \\
05.10.Gg;
 stochastic analysis methods   (Fokker-Planck, Langevin, etc); \\
02.50.Fz;
                stochastic processes; \\
%% 05.40.-a fluctuation phenomena, random processes, noise,
%%     and Brownian motion
%% 87.80.Vt Dynamical, regulatory, and integrative biology
%72.70.+m;
% Noise processes and phenomena;  \\
87.15.Ya
 Fluctuations

\end{abstract}

%\pacs{ PACS numbers:  ??? }

\maketitle

%\section{Formulation of the problems}

{\it One of the principle objects of theoretical research in any
department of knowledge is to find the point of view from which
the subject appears in its greatest simplicity.}

{ {\ }  \hspace{80mm} {\ } Josiah Willard Gibbs (1839-1903) }

\section{Introduction}

The theory proposed by Darwinian and Wallace
\cite{darwin1858,darwin1958} on the evolution in biology has been
the fundamental theoretical structure to understand biological
phenomena for nearly one and half centuries, referred to as the
Darwinian dynamics in the present paper. In its initial
formulation, the theory was completely narrative. No single
equation was used. There have been continuous effects to clarify
its meaning and to make it into more quantitative hence more
predictive \cite{fisher,wright,li,michod,waxman,ao2005}.
Tremendous progresses have made during past 100 years. Now the
degree of its usage of mathematics is comparable to any other
mathematically sophisticated natural science. From the physics
point of view, this theory is a {\it bona fide} nonequilibrium
dynamical theory.

In physics there has been a sustained interest during past several
decades in nonequilibrium processes
\cite{nicolis,keizer,vankampen,risken,cross,langer,gardiner}. The
important goals are to bridge its connection to equilibrium
processes and to clarify the roles of entropy and the Second Law
of thermodynamics. Thanks to recent progresses in experimental
technologies, particularly the nanotechnololgy, many previous
inaccessible regimes are now been actively explored. There have
renewed interests in this field, ranging from physics
\cite{lebowitz,gbg,physics,hanggi}, chemistry \cite{chemistry},
material science \cite{ottinger}, biology \cite{ao2005}, and to
many other fields \cite{feynmankac}. Quantitative experimental and
theoretical studies find their ways into the cellular and
molecular processes of life. There is a strong going interaction
between physical and biological sciences. The purpose of the
present paper is to look at the fundamental issues in statistical
mechanics and thermodynamics from the point of view of the
Darwinian dynamics and to gain a new insight.

There is even an active interest from philosophical point of view
on the foundation of statistical mechanics and thermodynamics.
Relevant to the present paper, following three fundamental but
controversial problems have been formulated \cite{uffink} : 1) In
what sense can thermodynamics said reduced to statistical
mechanics? 2) How can one derive equations that are not
time-reversal invariant from a time-reversal invariant dynamics?
3) How to provide a theoretical basis for the "approach to
equilibrium" or irreversible processes?

The Darwinian dynamics can answer all three questions in its own
way. For the first question, as long as the statistical mechanics
is formulated according to the Boltzmann-Gibbs distribution, the
main structures of statistical mechanics and thermodynamics are
equivalent. For the second question, it is found that the
thermodynamics is based on the energy conservation and on the
Carnot heat engine. It deals with quantities at equilibrium or
steady state without time. There is no direction of time. Hence,
there is no conflict between the thermodynamics and the
time-reversal dynamics. For the last and third question, Darwinian
dynamics comes with an adaptive behavior
\cite{darwin1858,darwin1958,fisher,wright,michod,waxman,ao2005}
and with a built-in direction of time. It naturally provides a
framework to address the question of "approaching to equilibrium".
If one would insist, the third question might be transformed into
another one: What would be the implication that there is a mutual
reduction between the Darwinian dynamics and the Newtonian
dynamics \cite{ao2005}?  Answer to this last question will not be
attempted in the present paper. The base to answer above three
questions will be discussed in next few sections.

The rest of the paper is organized as follows. In section II the
Darwinian dynamics will be summarized in the light of recent
progress. In section III it will be shown that the statistical
mechanics and canonical ensemble follows naturally from the
Darwinian dynamics. In section IV the connection to thermodynamics
is explored. There it will be shown that the Zeroth Law, the First
Law, and the Second Law can follow from the Darwinian dynamics,
not the Third Law. In section V two types of simple but seemly
profound dynamical equalities discovered recently, one based on
the Feynman-Kac formula and one a generalization of the Einstein
relation, are discussed. In section VI the present demonstration
is put into perspective.  No mathematical rigor is pursued in the
present paper, but the care has been taken to make the
demonstrations as clear as possible. With the solid physical and
biological foundations behind, a rigorous mathematical formulation
is possible.

%universal (thermodynamics) vs open (randomness)

\section{ Darwinian Dynamics, Adaptive Landscape, and F-Theorem }

This section summarizes the recent results on the Darwinian
dynamics.

\subsection{ Stochastic differential equation: the trajectory view }

In the context of genetics the Darwin's theory of evolution
\cite{darwin1858, darwin1958} may be summarized verbally  as that
the evolution is a result of genetic variation and its ordering
through elimination and selection. Both randomness and selection
are equally important in this dynamical process. With an suitable
time scale, the Darwinian dynamics may be represented by the
following stochastic differential equation \cite{li,ao2005}
\begin{equation}  \label{standard}
  \dot{{\bf q}} = {\bf f}({\bf q}) + N_I({\bf q}) {\bf \xi}(t) \; ,
\end{equation}
where ${\bf f}$ and ${\bf q}$ are $n$-dimensional vectors and
${\bf f}$ a nonlinear function of ${\bf q}$. The genetic frequency
of $i$-th trait is represented by $q_i$. Nevertheless, in the
present paper it will be treated as a generic real function of
time $t$. All quantities in this paper are dimensionless. They are
assumed to be measured in their own proper units unless explicitly
specified. The collection of all ${\bf q}$ forms a real
$n$-dimensional phase space. The noise ${\bf \xi}$ is a standard
Gaussian white noise with $l$ independent components: $
 \langle  \xi_i \rangle_{\xi} = 0 \; ,
$
and
\begin{equation}
 \langle \xi_i(t) \xi_j (t')\rangle_{\xi}
  = \theta \; \delta_{ij} \delta (t-t') \; ,
\end{equation}
and $i,j=1, 2, ..., l$. Here $ \langle ... \rangle_{\xi} $ denotes
the average over the noise variable $\{{\bf \xi} (t) \}$, to be
distinguished from the average over the distribution in phage
space below. The positive numerical constant $\theta$ describes
the strength of noise.

A further description of the noise term in Eq.(\ref{standard}) is
through the $n\times n$ diffusion matrix $D({\bf q})$, which is
defined by the following matrix equation
\begin{equation}
  N_I({\bf q}) N_I^\tau ({\bf q}) = 2 D({\bf q}) \; ,
\end{equation}
where $N_I$ is an $n\times l$ matrix,  $N_I^\tau$ is its the
transpose, which describes how the system is coupled to the noisy
source. This is the first type of the F-theorem \cite{ao2005}, a
generalization of Fisher's fundamental theorem of natural
selection \cite{fisher} in population genetics. According to
Eq.(2) the $n\times n$ diffusion matrix $D$ is both symmetric and
nonnegative. For the dynamics of state vector ${\bf q}$, all what
needed from the noisy term in Eq.(1) are the diffusion matrix $D$
and the positive numerical parameter $\theta$. Hence, it is not
necessary to require the dimension of the stochastic vector $\xi$
be the same as that of the state vector ${\bf q}$. This implies
that in general $l \neq n$.

It is known that a large class of nonequilibrium processes can be
described by such a stochastic differential equation
\cite{nicolis,keizer,vankampen,risken,cross,langer,gardiner}.
There is a strong current interest on such stochastic and
probability description ranging from physics
\cite{physics,hanggi}, chemistry, \cite{chemistry}, material
science \cite{ottinger}, biology \cite{ao2005}, and other fields
\cite{feynmankac}.

The Darwinian dynamics was conceived graphically by Wright in 1932
as the motion of the system in an adaptive landscape
\cite{wright,michod,waxman}. Since then such a landscape has been
known as the fitness landscape in some part of literature.
However, there are a considerable amount of confusion on the
definitions of fitness \cite{michod,ao2005}. In this paper a more
neutral term, the (Wright evolutionary) potential function, will
be used to denote this landscape. The adaptive landscape
connecting both the individual dynamics and its final destination
is intuitively appealing. Nevertheless, it had been difficult to
prove its existence in a general setting. The difficulty lies in
the fact fact that typically the detailed balance condition does
not hold in Darwinian dynamics, that is, $D^{-1}({\bf q}) {\bf
f}({\bf q})$ cannot be written as a gradient of scalar function
\cite{nicolis,vankampen,risken,cross,gardiner}.

{\ }

{\ }

{\ }

{\ }

 Figure 1. Adaptive landscape with in potential contour representation.
 $+$: local basin; $-$: local peak;
 $\times $: pass (saddle point).

During the study of the robustness of the genetic switch in a
living organism \cite{zhu} a constructive method was discovered to
overcome this difficulty: Eq.(1) can be transformed into the
following form of stochastic differential equation,
\begin{equation} \label{normal}
  [ R({\bf q}) + T({\bf q})] \dot{{\bf q}} = - \nabla
   \phi({\bf q}; \lambda) + N_{II}({\bf q})\xi(t) \; ,
\end{equation}
where the noise $\xi$ is from the same source as that in Eq.(1).
The parameter $\lambda$ denotes the influence of non-dynamical and
external quantities.  It should be pointed out that the potential
function $\phi$ may also implicitly depends on $\theta$. The
friction matrix $R({\bf q})$ is defined through the following
matrix equation
\begin{equation}
  N_{II}({\bf q}) N_{II}^\tau ({\bf q})= 2 R({\bf q}) \; ,
\end{equation}
which guarantees that $ R $ is both symmetric and nonnegative.
This is the second type of the F-theorem \cite{ao2005}. The
F-theorem emphasizes the connection between the adaption and
variation and is a reformulation of fluctuation-dissipation
theorem in physics \cite{kubo,zwanzig}. For simplicity we will
assume $\det( R ) \neq 0$ in the rest of the paper. Hence $\det( R
+ T ) \neq 0$ \cite{kat}. The breakdown of detailed balance
condition or the time reversal symmetry is represented by the
finiteness of the transverse matrix, $T \neq 0$. The usefulness of
the formulation of Eq.(\ref{normal}) is already manifested in the
successful solution of outstanding stable puzzle in gene
regulatory dynamics \cite{zhu} and in a consistent formulation of
the Darwinian dynamics \cite{ao2005}.

The $n\times n$ symmetric non-negative friction matrix $ R $ and
the transverse matrix $T$ are related to the diffusion matrix $D$:
$$
%\begin{equation}
 R({\bf q}) + T({\bf q})
  = {1 \over { D({\bf q}) + A({\bf q})} } \; .
%\end{equation}
$$
Here $A$ is an antisymmetric matrix determined by both the
diffusion matrix $D({\bf q})$ and the deterministic force ${\bf
f}({\bf q})$ \cite{kat,ao2004}. One of more suggestive forms of
above equation is
\begin{equation} \label{einstein}
 [ R({\bf q}) + T({\bf q}) ] D [ R({\bf q}) - T({\bf q}) ]
   = R({\bf q}) \; .
\end{equation}
This symmetric matrix equation implies $n(n+1)/2$ single equations
from each of its element. The Wright evolutionary potential
function $\phi({\bf q})$ is connected to the deterministic force
${\bf f}({\bf q})$ by
$$
%\begin{equation}
 - \nabla \phi({\bf q}; \lambda)
   =  [ R({\bf q}) + T({\bf q}) ] {\bf f}({\bf q}) \; .
%\end{equation}
$$
Or its equivalent form,
\begin{equation} \label{p-condition}
 \nabla \times [ [R({\bf q}) + T({\bf q}) ] {\bf f}({\bf q}) ] = 0 \; .
\end{equation}
Here the operation $\nabla \times$ on an arbitrary $n$-dimensional
vector ${\bf v}$ is a matrix generalization of the curl operation
in lower dimensions ($n=2,3$): $
 (\nabla \times {\bf v} )_{i,j} = \nabla _i  v_j
- \nabla_j  v_i \; . $ Above matrix equation is hence
antisymmetric and gives $n(n-1)/2$ single equations from each of
its element. From Eq.(\ref{einstein}) and (\ref{p-condition}) the
friction matrix $R$, the transverse matrix $T$, and the potential
function $\phi$ can be constructed in terms of the diffusion
matrix $D$ and the deterministic force ${\bf f}$. The local
construction was demonstrated in detail in Ref.\cite{kat}. For a
global construction an iterative method was outlined in Ref.
\cite{ao2004} .

In the case the stochastic drive may be ignored, that is, $\theta
= 0 $, the relationship between Eq.(1) and (\ref{normal}) remains
unchanged. Furthermore, Eq.(\ref{normal}) becomes a deterministic
equation
\begin{equation} \label{adaptive}
  [ R({\bf q}) + T({\bf q})] \dot{{\bf q}} = - \nabla
   \phi({\bf q}; \lambda)  \; .
\end{equation}
Because of the non-negativeness of the friction matrix, one
obtains
\begin{eqnarray}
 \frac{d }{d t} \phi({\bf q}; \lambda)
   & = & \dot{{\bf q}} \cdot  \nabla \phi({\bf q}; \lambda)
   \nonumber \\
   & = & - \dot{{\bf q}}^{\tau}  [ R({\bf q}) + T({\bf q}) ]
    \dot{{\bf q}} \nonumber \\
   & = & - \dot{{\bf q}}^{\tau}  R({\bf q}) \dot{{\bf q}}
   \nonumber \\
   & \leq &  0 \; .
\end{eqnarray}
It is immediately clear that the Wright evolutionary potential
function $\phi({\bf q}; \lambda)$ is a Lyapunov function and the
deterministic dynamics makes it non-increasing: the tendency to
approach the nearby potential minimum to achieve the maximum
probability. This is precisely what conceived by Wright. The
adaptive dynamics has been actively exploring in biology
\cite{waxman}.

The conservative Newtonian dynamics may be regarded as a further
limit of zero friction matrix, $ R=0 $. Hence, from
Eq.(\ref{adaptive}), the Newtonian dynamics may be expressed as,
\begin{equation} \label{newton}
  T({\bf q}) \; \dot{{\bf q}} = - \nabla  \phi({\bf q}; \lambda)  \; .
\end{equation}
Here the value of potential function is evidently conserved during
the dynamics:  $\dot{{\bf q}} \cdot  \nabla \phi({\bf q}; \lambda)
= 0 $, that is, the system moves along the equal potential contour
in the adaptive landscape.

\subsection{Fokker-Planck equation: the ensemble view }

It was heuristically argued \cite{ao2004} that the steady state
distribution $\rho({\bf q})$ in the state space is, if exists,
\begin{equation} \label{bg}
 \rho({\bf q},t=\infty) \propto
  e^{ - \beta {\phi({\bf q}; \lambda ) } } \; .
\end{equation}
Here $\beta = 1/\theta$. It takes the form of Boltzmann-Gibbs
distribution function. Therefore, the potential function $\phi$
acquires both the dynamical meaning through Eq.(\ref{normal}) and
the steady state meaning through Eq.(\ref{bg}).

It was further demonstrated that such a heuristical argument can
be translated into an explicit procedure such that there is an
explicit Fokker-Planck equation whose steady state solution is
indeed given by Eq.(\ref{bg}) \cite{ya}. Starting for the the
generalized Klein-Kramers equation, taking the limiting procedure
of the zero mass limit, the desired Fokker-Planck equation
corresponding to Eq.(\ref{normal}) is
\begin{equation} \label{fp-eq}
 {\partial \rho({\bf q},t) \over \partial t}
  = \nabla^{\tau} [D({\bf q}) + A({\bf q}) ] [\theta \nabla
   + \nabla \phi({\bf q}; \lambda)] \rho({\bf q},t) \; .
\end{equation}
This equation is also a statement of conservation of probability.
It can be rewritten as the probability continuity equation:
\begin{equation} \label{continuity}
 {\partial \rho({\bf q},t) \over \partial t}
  + \nabla \cdot {\bf j}({\bf q}, t) = 0  \, ,
\end{equation}
with the probability current density ${\bf j}$
\begin{equation} \label{current}
 {\bf j}({\bf q}, t) \equiv - [D({\bf q}) + A({\bf q}) ]
   [\theta \nabla + \nabla \phi({\bf q}; \lambda)]
   \rho({\bf q},t) \; .
\end{equation}
The reduction of dynamical variables has often been done by the
well-known Smoluchowski limit. In the above derivation we take the
mass to be zero, keeping other parameters, including the friction
and transverse matrices, to be finite. Nevertheless, in the
Smoluchowski limit it is the friction matrix to be taken as
infinite, keep all other parameters to be finite. Those two limits
are in general not exchangeable.

The steady state configuration solution of Eq.(\ref{fp-eq}) is
indeed given by Eq.(\ref{bg}). It would be interested to point out
that the steady state distribution function, Eq.(\ref{bg}), is
independent of both friction matrix $ R $ and the transverse
matrix $ T $. Furthermore, we emphasize that no detailed balance
condition is assumed in reaching this result. In addition, both
the additive and multiplicative noises are treated here on equal
footing.

Finally, it can be verified that above construction leading to
Eq.(\ref{fp-eq}) is valid and remains unchanged when there is an
explicit time dependent in $R$, $T$, and/or $\phi$. In this case
though there may not exist a steady state distribution if the
Wright evolutionary potential function $\phi$ is time dependent.

\section{ Statistical Mechanics }

\subsection{Central Relations in Statistical Mechanics}

As discussed above, if treating the parameter $\theta$ as
temperature, the steady state distribution function in phase space
is indeed the familiar Boltzmann-Gibbs distribution,
Eq.(\ref{bg}). The partition function, or the normalization
constant, is then
\begin{equation} \label{p-function}
 {\cal Z}_{\theta}(\lambda) \equiv \int d {\bf q} \,
  e^{ - \beta {\phi({\bf q}; \lambda) } } \; .
\end{equation}
The integral $\int d {\bf q}$ denotes the summation over whole
phase space. The normalized steady state distribution is
\begin{equation}
 \rho_{\theta}({\bf q})
  \equiv  \frac{e^{ - \beta {\phi({\bf q}; \lambda) } } }
   { {\cal Z}_{\theta} } \; .
\end{equation}
For a given observable quantity $O({\bf q})$, its average or
expectation value is
\begin{eqnarray} \label{summit}
 \langle O \rangle_{\bf q}
  & \equiv &  \int d {\bf q} \, O({\bf q}) \,
       \rho_{\theta}({\bf q}) \nonumber \\
  & = &  \frac{1}{{\cal Z}_\theta} \int d {\bf q} \, O({\bf q}) \,
  e^{ - \beta {\phi({q}; \lambda) } } \: .
\end{eqnarray}
The subscript $q$ denoted that the average is over phage space,
not over the noise in Eq.(\ref{standard}) or (\ref{normal}).
Eq.(\ref{summit}) is the summit of statistical mechanics.

\subsection{Stochastic process and canonical ensemble}

A main question is that for a given Fokker-Planck equation, can
the corresponding stochastic differential equation in the form of
Eq.(\ref{normal}) be recovered? The answer is affirmative and the
procedure to carry it out is already contained in
Eq.(\ref{fp-eq}), which will be briefly demonstrated below.

A generic form for the Fokker-Planck equation may be expressed as
follows:
\begin{equation} \label{fp-generic}
  {\partial \rho({\bf q},t) \over \partial t}
  = \nabla^{\tau} [\theta  \overline{D}({\bf q}) \nabla
    - \overline{{\bf f}}({\bf q})] \rho({\bf q},t) \; .
\end{equation}
Here ${\overline{D}}({\bf q})$ is the diffusion matrix and
$\overline{{\bf f}}({\bf q})$ the drift force. The main motivation
to take such a form is simple: In the case detailed balance
condition is satisfied, {\it i.e.}, $A({\bf q}) = 0$ (and $T({\bf
q}) = 0$), the potential function $\overline{\phi}$ can be
directly read from above equation: $\nabla \overline{\phi} =
\overline{D}^{-1} \; \overline{\bf f}$. It puts the diffusion
effect in a very prominent position. Any other form of
Fokker-Planck equations can be easily transformed into above form.
This generic form of the Fokker-Planck equation is less tangible
to additional complications such as the noise induced first order
transitions caused by the ${\bf q}$-dependent diffusion constant.

A potential function $\overline{\phi}({\bf q})$ can always be
defined from the steady state distribution. There is an extensive
mathematical literature addressing this problem \cite{doob}. After
this is done, though it can be a difficult mathematical problem,
the procedure to relate the genetic Fokker-Planck equation to
Eq.(\ref{fp-eq}) is particularly straightforward. Eq.(\ref{fp-eq})
can be rewritten as
\begin{equation}  \label{fp-eq2}
 {\partial \rho({\bf q},t) \over \partial t}
  = \nabla^{\tau} [\theta D({\bf q})\nabla
   + \theta (\nabla^{\tau} A({\bf q}))
   -  [D({\bf q}) + A({\bf q})]\nabla \phi({\bf q})]
      \rho({\bf q},t) \; .
\end{equation}
The antisymmetric property of the matrix $A({\bf q})$ has been
used in reaching Eq.({\ref{fp-eq2}).  Thus, comparing between
Eq.(\ref{fp-generic}) and (\ref{fp-eq2}), we have
\begin{eqnarray}
  D({\bf q})
   & = & \overline{D}({\bf q}) \; , \\
  \phi({\bf q})
   & = & \overline{\phi}({\bf q}) \; , \\
  {\bf f}({{\bf q}})
   & = & \overline{{\bf f}}({\bf q})
    + \theta \nabla^{\tau} A({\bf q}) \; . \label{force}
\end{eqnarray}
In reaching Eq.(\ref{force}) we have used the relation
$$
 - [D({\bf q}) + A({\bf q})] \nabla \phi({\bf q})
   = {\bf f}({\bf q}) \; .
$$
The explicit equation for the anti-symmetric matrix $A$ is
\begin{equation} \label{Q-eq}
  \theta \nabla^{\tau} A({\bf q})
   + [ D({\bf q}) + A({\bf q}) ]  \nabla \phi({\bf q}; \lambda)
    =  \overline{{\bf f}}({\bf q}) \; ,
\end{equation}
which is a first order linear inhomogeneous partial differential
equation. The solution for $A$ can be formally written down
\begin{equation}
  A({\bf q}) =  {1 \over{\theta}} \int^{\bf q} d{\bf q}'
   [ \overline{\bf f}({\bf q}')
    - D({\bf q}')\nabla ' \phi({\bf q}'; \lambda) ]
    e^{ \beta ({\phi({\bf q}; \lambda)  - \phi({\bf q}'; \lambda)} ) }
  + A_0({\bf q}) e^{  \beta \phi({\bf q}; \lambda ) }  \; .
\end{equation}
Here $A_0({\bf q})$ is a solution of the homogenous equation
$\theta \nabla^{\tau} A({\bf q}) = 0 $ and the two parallel
vectors in the integrand, such as $d{\bf q}' \; \overline{\bf
f}({\bf q})$, forms a matrix. This completes our answer to the
converse question.

It is interesting to note that the shift between the zero's of the
potential gradient and the drift is given by, from
Eq.(\ref{force}),
\begin{equation} \label{shift}
 \Delta \overline{\bf f}
  = \theta \nabla^{\tau} A({\bf q}) \; ,
\end{equation}
that is, the extremals of the steady state distribution are not
necessary determined by the zero's of drift. This is the formula
for such a shift shown extensively in numerical studies
\cite{shift}. This shift can occur even when $D=constant$.

Thus, the zero-mass limit approach to the stochastic differential
equation is consistent in itself. The meaning of the potential
$\phi$ is explicitly manifested in both local trajectory according
to Eq.(\ref{normal}) and ensemble distribution according to
Eq.(\ref{fp-eq}). In particular, no detailed balance condition is
assumed. There is no need to differentiate between the additive
and multiplicative noises. This zero mass limit procedure which
leads to Eq.(\ref{normal}) from Eq.(\ref{fp-eq}) may be regarded
as another prescription for the stochastic integration, in
addition to those of Ito and Stratonovich
\cite{vankampen,risken,gardiner}. The connection to those methods
of treating stochastic differential equation is suggested through
Eqs.(\ref{fp-generic}) and (\ref{fp-eq}) (or Eq.(\ref{fp-eq2}) ).
The Ito's method puts an emphasis on the martingale property of
stochastic processes, which may be viewed as a prescription from
mathematics. The Stronotovich method stresses the
differentiability such that the usual differential chain-rule can
be formally applied, which may be viewed as the prescription from
engineering. The present approach emphasizes the role played by
the potential function in both trajectory and ensemble
descriptions. It may be regarded as the prescription from natural
sciences.

We may conclude that the stochastic process, regardless of Ito,
Stratonovich, the present method, or others, leads to the
canonical ensemble with a temperature and a Boltzmann-Gibbs type
distribution function.

\subsection{Discrete stochastic dynamics}

There is another kind of modelling predominant in population
genetics and other fields which is discrete in phage space and/or
time. Here we would not get into it in any detail, except quoting
results when necessary. The reasons of being able to do so are: 1)
It is known mathematically any discrete model can be represented
by a continuous one exactly, though sometimes such a process may
turn a finite dimension problem into an infinite dimension one; 2)
By a coarse graining average the discrete dynamics in population
genetics can often be simplified to continuous ones such as
diffusion equations or Fokker-Planck equations
\cite{li,vankampen,gardiner}. It is generally acknowledged in
population genetics and in other fields that the diffusion
approximation is a good start and usually accurate.

For the steady state distribution, all one needs to know is the
Wright evolutionary potential function $\phi$ and the positive
numerical constant $\theta$ which in many cases can be set to be
unity: $\theta = 1$. Hence, discrete or continuous representation
is not a physically or biologically relevant point.

\section{ Thermodynamics }

Given the Boltzmann-Gibbs distribution, the partition function can
be evaluated according to Eq.(\ref{p-function}). Hence, at the
steady state, all observable quantities are in principle known
according to Eq.(\ref{summit}). One may wonder then what would be
the value of thermodynamics. First, there is a practical reason.
In many cases the calculation of the partition function is a hard
problem, if possible. It would be desirable if there are
alternatives. Thermodynamics gives us a set of useful relations
between observable quantities based on general properties of the
system such as symmetries. Useful and precise information on one
phenomenon can be inferred from the information on other
quantities. Second, there is a theoretical reason. The
thermodynamics has a scope far more general than most other fields
in physics. It is the only field in classical physics whose
foundation and structure not only have survived quantum mechanics
and relativity shakeups, but become stronger. Furthermore,
thermodynamics has a formal elegance which is exceedingly
satisfying aesthetically. Its influence is far beyond physical
sciences.

There exists already numerous excellent books exposing the
thermodynamics from statistical mechanics point of view. A
thorough treatment can be found in Callen \cite{callen}. A
reader-friendly treatment can be found in Ma \cite{ma}. Concise
and elementary treatments from thermodynamics point of view were
given by Pippard \cite{pippard} and by Reiss \cite{reiss}. In the
light of those superb expositions, the present discussion may
appear incomplete as well as arbitrary. For a systematic
discussion on thermodynamics the reader is sincerely encouraged to
consult those books and/or any of her/his favorites not listed
here. The main objective here is to show that the Darwinian
dynamics indeed implies the main structure of thermodynamics,
though at a first glance it seems to have no connection. The
Darwinian dynamics is at the extreme end of nonequilibrium
processes.

Even given a limited scope of presentation there are already
excellent and recent papers. The paper by Oono and Panconi
\cite{oono} is such an example. It gave a comprehensive review on
the problems from the point of view steady state thermodynamics.
There are overlaps at various places. Nevertheless, there is one
main difference: The role of "temperature" is emphasized here,
instead. The paper by Sekimoto \cite{sekimoto} gave a detailed
discussion on the connection between thermodynamics and Langevin
dynamics. The main difference is that in the present paper the
detailed balance condition is not needed.

\subsection{Zeroth Law}

From the Darwinian dynamics, the steady state distribution is
given by a Boltzmann-Gibbs type distribution, Eq.(\ref{bg}),
determined by the Wright evolutionary potential function $\phi$ of
the system and a positive parameter $\theta$ of the noise
strength. Hence, the analogy of the Zeroth Law of thermodynamics
is implied by the Darwinian dynamics: There exists a
temperature-like quantity, represented by the positive parameter
$\theta$. This "temperature" $\theta$ is absolute in that it does
not depend on the system's material details.

\subsection{First Law}

From the partition function ${\cal Z}_{\theta}$, we may define a
quantity
\begin{equation} \label{f-energy}
 F_{\theta} \equiv - \theta \ln {\cal Z}_{\theta} \; .
\end{equation}
We may also define the average Wright evolutionary potential
function,
\begin{equation} \label{i-energy}
 U_{\theta} \equiv \int d {\bf q} \; \phi({\bf q}; \lambda) \;
   \rho_{\theta}({\bf q}) \; .
\end{equation}
From the distribution function we may further define a positive
quantity
\begin{equation} \label{entropy}
 S_{\theta} \equiv - \int d {\bf q} \;
  \rho_{\theta}({\bf q}) \ln \rho_{\theta}({\bf q})  \; .
\end{equation}
It is then straightforward to verify that
\begin{equation}\label{first}
 F_{\theta}= U_{\theta} - \theta \; S_{\theta} \; ,
\end{equation}
precisely the fundamental relation in thermodynamics satisfied by
free energy, internal energy, and entropy. Hence we have the free
energy $F_{\theta}$, the internal energy $U_{\theta}$, and the
entropy $S_{\theta}$. The subscript $\theta$ emphasizes the steady
state nature of those quantities. Due to the finite strength of
stochasticity, that is, $\theta > 0$, not all $U_{\theta}$ is
readily usable: $F_{\theta}$ is always smaller than $U_{\theta}$.
A part of $ \theta \; S_{\theta}$ called "heat" cannot be
utilized.

It can also be verified from definitions that if the system
consists of several non-interacting parts, $F_{\theta}$,
$U_{\theta}$, and $S_{\theta}$ are sum of those corresponding
parts. Hence, they are extensive quantities. No attention is paid
here to the fine difference between additive and extensive
properties. Instead, the "temperature" ${\theta}$ is an intensive
quantity: it must be the same for all those parts because they are
contacting the same noisy source. Therefore, we conclude that the
analogy of the First Law of thermodynamics is implied in the
Darwinian dynamics.

The fundamental relation for the free energy, Eq.(\ref{first}), as
well as the internal energy, Eq.(\ref{i-energy}), may be expressed
in their differential forms. Considering an infinitesimal process
which causes changes in both the Wright evolutionary potential
function via parameter $\lambda$ and in the steady state
distribution function, the change in the internal energy according
to Eq.(\ref{i-energy}) is
\begin{eqnarray} \label{d-form1}
 d U_{\theta} & = & \int d {\bf q} \;
   \frac{ \phi({\bf q}; \lambda) }{\partial \lambda} \; d
   \lambda \;  \rho_{\theta}({\bf q})
    + \int d {\bf q} \; \phi({\bf q}; \lambda) \;
      d \rho_{\theta}({\bf q}) \nonumber \\
    & = & \mu \; d \lambda  +  \theta d S_{\theta}  \; .
\end{eqnarray}
This is the differential form for the internal energy. Here the
steady state entropy definition of Eq.(\ref{entropy}) has been
used, along with $ \int d {\bf q} \; d \rho_{\theta}({\bf q}) =
0$,  and
\begin{equation}
 \mu \equiv \left. \frac{ \partial U_{\theta} }{\partial \lambda } \right|
  _{\theta}  \; .
\end{equation}
Eq.(\ref{d-form1}) can be written in the usual form in
thermodynamics:
$$
  d U_{\theta} = {\bar d} W + {\bar d}Q \; .
$$
The part corresponding to the change in entropy is the "heat"
exchange: ${\bar d}Q = \theta \; d S$ and the part corresponding
to the change in the Wright evolutionary potential function is the
"work" $ {\bar d} W = \mu \; d \lambda$. The conservation of
"energy" is most clearly represented by Eq.(\ref{d-form1}). For
the free energy,
\begin{eqnarray} \label{d-form2}
 d F_{\theta}
  & = & d U_{\theta}  - d \theta \; S_{\theta} - \theta d S_{\theta} \nonumber \\
  & = & \mu \; d \lambda - S_{\theta} \;  d \theta  \, .
\end{eqnarray}
Eq.(\ref{d-form1}) and (\ref{d-form2}) may be useful in some
applications. For example, the "temperature" can be found via
Eq.(\ref{d-form1}):
\begin{equation} \label{temperature}
 \theta = \left. \frac{\partial U_{\theta}}
              {\partial S_{\theta}} \right| _{\lambda} \, .
\end{equation}
This relation may be used to find the "temperature" in a
nonequilibrium process if it is not obvious to identify {\it a
priori} \cite{jou}.

The convexity of a thermodynamical quantity is naturally
incorporated by the Boltzmann-Gibbs distribution. There is no
restriction on the size of the system. Even for a finite system,
phase transitions can occur, because singular behaviors can be
built into the potential function, and controlled by external
quantities.

\subsection{Second Law}

First, we remind the reader of a few more important definitions. \\
A {\it reversible process} is such a process that all the relation
between quantities and parameters in question is defined through
the Boltzmann-Gibbs distribution, Eq.(\ref{bg}). From the
Darwinian dynamics point of view, the reversible process in
reality is necessarily a slow or quasi-static process in order to
ensure the relevancy of steady state distribution
for its any practical realization. \\
An {\it isothermal process} is a reversible process in which
"temperature" $\theta$ remains unchanged, $\theta = constant$. No
confusion with the thermostated processes, which are in general
nonequilibrium dynamical processes, should arises. \\
An {\it adiabatic process} is a reversible process in which the
coupling between the system and the noise source is switched off
and the system vary in such a way the distribution function
remains unchanged along the dynamics trajectory when following
each point in phase space. This implies that the entropy remains
unchanged, $S_{\theta} = constant$. The adiabatic process has
often been used in irreversible processes in that there is no heat
exchange between the system and the noisy environment, hence $S(t)
= constant$ ({\it c.f.} Eq.({\ref{g-entropy})).

Now we discuss the analogy of Carnot cycle on which the Carnot
heat engine is based. The Carnot cycle consists of four reversible
processes: two isothermal processes and two adiabatic processes
(Fig. 1.a,b). The efficiency $\nu$ of the Carnot heat engine is
defined as the ratio of the total net work performed over the heat
absorbed at high temperature:
\begin{equation}
 \nu \equiv \frac{ \Delta W_{total} } {\Delta Q_{12} } \; .
\end{equation}

 {\ }

 {\ }

 {\ }

 {\ }

 \hspace{50mm}  {\ } (a) \hspace{50mm} (b)

 Figure 2. Carnot cycle. (a). The $\mu-\lambda$ representation.
 (b). The $\theta-S$ representation. In this temperature-entropy
 representation, the Carnot cylce is a rectangular.

The total net work done by the system is represented by the shaded
area enclosed by the cycle. For the heat absorbed at the high
isothermal process $1 \rightarrow 2 $,
\begin{equation}
 \Delta Q_{12} = \theta_{high} \Delta S_{\theta,12}  \; .
\end{equation}
For the adiabatic process $2 \rightarrow 3$, an external
constraint represented by $\lambda$ is released (or applied),
\begin{equation}
 \Delta S_{\theta,23} = 0 {\ } ,  {\ } \Delta Q_{23} = 0  \; .
\end{equation}
For the heat absorbed (rather, released) at the low isothermal
process $3 \rightarrow 4$,
\begin{equation}
 \Delta Q_{34} = \theta_{low} \Delta S_{\theta,34} = - \Delta Q_{43}  \; .
\end{equation}
For the adiabatic process $4 \rightarrow 1$, an external
constraint is applied (released),
\begin{equation}
 \Delta S_{\theta,41} = 0 {\ } , {\ }  \Delta Q_{41} = 0 \; .
\end{equation}
Using the First Law, Eq.(\ref{first}) and the fact that the free
energy is  a state function
\begin{eqnarray}
 \Delta F_{total}
  & = & \Delta Q_{total} - \Delta W_{total}  \nonumber \\
  & = &  0  \; .
\end{eqnarray}
The minus sign in front of the total work represents that it is
the work done by the system, not to the system. The total heat
absorbed by the system is
$$
 \Delta Q_{total} =  \Delta Q_{12} + \Delta Q_{34}
   =   \Delta Q_{12} - \Delta Q_{43} = \Delta W_{total} \, .
$$
We further have
\begin{equation}
 \Delta S_{\theta,12} = \Delta S_{\theta,43} \; .
\end{equation}
From Eq.( ), ( ) and ( ) the Carnot heat engine efficiency is then
\begin{eqnarray} \label{efficiency}
 \nu & = & 1 - \frac{ \Delta Q_{43}}{\Delta Q_{12} }  \nonumber \\
     & = & 1 - \frac{ \theta_{low}}{\theta_{high} }    \; ,
\end{eqnarray}
precisely the form in thermodynamics. The beauty of Carnot heat
engine is that its efficiency is completely independent of any
material details. It brings out the most fundamental property of
thermodynamics and is a direct consequence of the Boltzmann-Gibbs
distribution function and the First Law. It reveals a property of
Nature which may not be contained in a conservative dynamics, at
least it is still not obviously to many people from the Newtonian
dynamics point after more than 150 years of intensive studies. The
Second Law of thermodynamics may be stated as that for all heat
engines operating between two temperatures, Carnot heat engine is
the most efficient. The Second Law is implied in the Darwinian
dynamics.

There are many other versions of the Second Law, on which the
reader is suggested to consult the books listed at the beginning
of this section. Here we mentioned two equivalent
versions from the stability point of view, which frame following discussions. \\
Minimum free energy statement: For given the potential function
and the temperature, the free energy achieves its lowest possible
value if the distribution is the Boltzmann-Gibbs
distribution. \\
Maximum entropy statement: For given potential function and its
average, the entropy attains its maximum value when the
distribution is the Boltzmann-Gibbs distribution. This version of
the Second Law is the most influential. Its inverse statement, the
so-called maximum entropy principle, has been extensively employed
in the probability inference \cite{jaynes} both within and beyond
physical and biological sciences.

It is attempting to generalize the entropy definition to the
arbitrary time dependent distribution in analogy to
Eq.(\ref{entropy}):
\begin{equation} \label{g-entropy}
 S(t) \equiv
  - \int d {\bf q} \; \rho({\bf q}, t) \; \ln \rho({\bf q}, t) \; .
\end{equation}
There are two apparent drawbacks for such definition, however.
First, even if the evolution of the distribution function $
\rho({\bf q}, t) $ is governed by the Fokker-Planck equation,
Eq.(\ref{fp-eq}), in general the sign of its time derivative, $ d
S(t)/d t = {\dot S}(t)$, cannot be determined, whether or not it
is close to the steady state distribution. Though $ {\dot S}(t) $
might indeed be divided into an always positive part and the rest,
such a partition is arbitrary. More seriously, in general $S(t)$
can be either larger or smaller than $S_{\theta}$, which makes
such a definition lose its appealing in the view of the maximum
entropy statement of the Second Law. We will return to $S(t)$
later.

Nevertheless, if taking the lesson from the potential function
that only the relative value is important, we may introduce a
reference point in the functional space into a general entropy
definition. One definition for the referenced entropy is
\cite{mazur}
\begin{equation} \label{r-entropy}
 S_r(t) \equiv
  - \int d {\bf q} \; \rho({\bf q}, t) \;
   \ln \frac{ \rho({\bf q}, t) }{\rho_{\theta}({\bf q}) }
   + S_{\theta} \; .
\end{equation}
With the aid of inequality $\ln(1+x) \leq x $ and the
normalization condition $ \int d {\bf q} \; \rho({\bf q}, t) =
\int d {\bf q} \; \rho_{\theta}({\bf q}) = 1$, it can be
immediately verified that
\begin{eqnarray}
  S_r(t)
   & = & \int d {\bf q} \; \rho({\bf q}, t) \;
    \ln \left( 1 + \frac{ \rho_{\theta}({\bf q}) - \rho({\bf q}, t) }
                  {\rho({\bf q}, t) } \right)
   + S_{\theta} \; ,   \nonumber \\
   & \leq & \int d {\bf q} \; ( \rho_{\theta}({\bf q})
            - \rho({\bf q}), t )
           + S_{\theta} \; , \nonumber \\
   & \leq &  S_{\theta}  \; .
\end{eqnarray}
The equality holds when $\rho({\bf q},t) = \rho_{\theta}({\bf
q})$. This inequality is independent of the details of the
dynamics and is evidently a maximum entropy statement.
Furthermore, with the aid of the Fokker-Planck equation,
Eq.(\ref{fp-eq}), the time derivative of this referenced entropy,
$d S_r(t)/d t = {\dot S}_r(t)$ is always non-negative:
\begin{eqnarray}
 {\dot S}_r(t)
  & = & - \int d {\bf q} \;
    \frac{\partial \rho}{\partial t}({\bf q}, t) \;
    \ln \frac{ \rho({\bf q}, t) }{\rho_{\theta}({\bf q}) }  \nonumber \\
  & = & - \int d {\bf q} \;
    \left( \nabla^{\tau} [D({\bf q}) + A({\bf q}) ] [\theta \nabla
   + \nabla \phi({\bf q}; \lambda)]  \rho({\bf q}, t) \right) \;
    \ln \frac{ \rho({\bf q}, t) }{\rho_{\theta}({\bf q}) } \nonumber \\
  & = &  \int d {\bf q} \;
     \left(\nabla \ln \frac{ \rho({\bf q}, t) }
                           {\rho_{\theta}({\bf q})} \right)^{\tau}
      [D({\bf q}) + A({\bf q}) ] [\theta \nabla
   + \nabla \phi({\bf q}; \lambda)] \rho({\bf q}, t) \nonumber \\
  & = &  \int d {\bf q} \;
     \frac{1}{\theta \rho({\bf q},t) }
     \left([\theta \nabla + \nabla \phi({\bf q}; \lambda)]
            \rho({\bf q}, t) \right)^{\tau}
      [D({\bf q}) + A({\bf q}) ]
       [\theta \nabla+ \nabla \phi({\bf q}; \lambda)]
           \rho({\bf q}, t)  \nonumber \\
  & = &  \int d {\bf q} \;
     \frac{1}{\theta \rho({\bf q},t) }
     \left( [\theta \nabla + \nabla \phi({\bf q}; \lambda)]
            \rho({\bf q}, t) \right)^{\tau}
      \; D({\bf q}) \;
       [\theta \nabla+ \nabla \phi({\bf q}; \lambda)]
           \rho({\bf q}, t)  \nonumber \\
  & = &  \int d {\bf q} \; \frac{1}{\theta \rho({\bf q},t) }
     {\bf j}^{\tau}({\bf q}, t) \; R({\bf q}) \; {\bf j}({\bf q}, t)
      \nonumber \\
   & \geq & 0    \; .
\end{eqnarray}
Hence, this referenced entropy $S_r(t)$ has all the desired
properties for the maximum entropy statement.

We remark that by the probability current density definition of
Eq.(\ref{current}) ${\bf j}$ is zero at the steady state. This may
differ from the usual probability current density definition which
may be based on Eq.(\ref{fp-generic}) and takes the form ${\bar
{\bf j}}({\bf q},t) \equiv - [\theta D \nabla - {\bar {\bf
f}}({\bf q})] \rho({\bf q}, t)$, which is not zero at the steady
state. Instead, $\nabla \cdot {\bar {\bf j}} = 0$ at the steady
state.

Though the general definition of entropy of Eq.(\ref{g-entropy})
may not be appealing, a general definition of free energy is
consistent with the Second Law. We demonstrate it here. First, a
general definition for the internal energy may be:
\begin{equation} \label{g-i-energy}
 U(t) \equiv  \int d {\bf q} \;
   \phi({\bf q}; \lambda) \; \rho({\bf q}, t) \; .
\end{equation}
Given the distribution and the potential function, quantities
defined in Eq.(\ref{g-entropy}) and (\ref{g-i-energy}) can be
evaluated. Following the form of Eq.(\ref{f-energy}) a general
definition of free energy would be, with the "temperature"
$\theta$,
\begin{equation} \label{g-f-energy}
 F(t) \equiv U(t) - \theta \; S(t) \; .
\end{equation}
It can be verified that $F(t) \geq F_{\theta}$ and its time
derivative is always non-positive, ${\dot F}(t) \leq 0 $. So
defined time dependent free energy indeed satisfies the minimum
free energy statement of the Second Law. It differs from the
referenced entropy $S_r(t)$ by a minus sign and by a constant:
$$
  F(t) = -\theta S_r(t) + U_{\theta} \; .
$$
The generalized entropy $S(t)$ has one desired property regarding
to the adiabatic processes (either reversible or irreversible) in
that $D=0$ during the adiabatic process. Hence,
\begin{eqnarray}
 {\dot S}(t)
  & = & - \int d {\bf q} \;
    \frac{\partial \rho}{\partial t}({\bf q}, t) \;
     \ln \rho({\bf q}, t)  \nonumber \\
  & = & - \int d {\bf q} \;
    \left[ \nabla^{\tau} \;  A({\bf q}) \;
      \nabla \phi({\bf q}; \lambda)  \rho({\bf q}, t) \right] \;
    \ln  \rho({\bf q}, t)  \nonumber \\
  & = & - \int d {\bf q} \;
    \left[(A({\bf q})\; \nabla \phi({\bf q}; \lambda)) \cdot \nabla \right] \;
     \int^{\rho({\bf q}, t)} d \rho ' \ln  \rho '  \nonumber \\
  & = &  0 \; .
\end{eqnarray}
This is the known result in conservative Newtonian dynamics that
the entropy remains unchanged. In deriving above equation we have
used two properties: 1) the no-coupling to the noisy environment
has been translated into the fact that the terms associated with
the diffusion matrix $D$ and "temperature" $\theta$ are set to be
zero in Eq.(\ref{fp-eq}), because they are related to the noisy
source whose information is not available during an adiabatic
process; and 2) the incompressible condition of $ \nabla \cdot [
A({\bf q}) \;
 \nabla \phi({\bf q}; \lambda) ] = 0 $, which is typically satisfied in
the Newtonian dynamics. In this conservative case, it can be
verified that ${\dot S}_r(t) = 0$, too, for any adiabatic process.

It may be worthwhile to mention another referenced entropy
$S_{r2}(t)$  which approaches the steady state entropy
$S_{\theta}$ from above. It's form is simple:
\begin{equation}
 S_{r2}(t) \equiv - \int d {\bf q} \;
    \rho_{\theta}({\bf q}) \; \ln \rho({\bf q}, t) \; .
\end{equation}
It can be verified that $S_{r2}(t) \geq S_{\theta}$ and ${\dot
S}_{r2}(t) \leq 0 $.

\subsection{Third Law}

Now we consider the behavior near zero "temperature", $\theta
\rightarrow 0$.  To be specific we assume the system is dominated
by a stable fixed point. As suggested by the Boltzmann-Gibbs
distribution, Eq.(\ref{bg}), only the regime of phase space near
this stable fixed point will be important. Hence the Wright
evolutionary potential function can be expanded around this point,
taking as ${\bf q} = 0$:
\begin{equation}
 \phi({\bf q}; \lambda) = \phi(0; \lambda) + \frac{1}{2}
  \sum_{j=1}^{n} k_j(\lambda) q_j^2  {\ } .
\end{equation}
Here we have also assumed that the number of independent modes is
the same as the dimension of the phase space, though it may not
necessary be so. This assumption will not affect our conclusion
below. Those independent modes are represented by $q_j$ without
loss of generality.  The "spring coefficients" $ \{ k_j \}$ are
functions of external parameters represented  by $\lambda$.

The partition function according to Eq.(\ref{p-function}) can be
readily evaluated in this situation:
\begin{equation}
 {\cal Z}_{\theta} = e^{-\beta \phi(0; \lambda)}  \; \prod_{j=1}^{n}
   \sqrt{\frac{2\pi \theta }{k_{j}} } {\  } .
\end{equation}
So is the entropy according according to Eq.(\ref{entropy}):
\begin{equation}
 S_{\theta} =  n  \left[ \theta -  \frac{1}{2} \ln\theta \right]
  + \frac{1}{2} \sum_{j}^{n} \frac{k_j}{2\pi}\, .
\end{equation}
The first term does not depend on external parameters, but the
second term does.  This suggests that the entropy depends on
control process in a finite manner at low enough temperature.
Hence, the Darwinian dynamics does not imply the Third Law in
which it states that in the limit of zero temperature the
difference in entropy between different processes is zero.

One should not be surprised by above conclusion, because the
Darwinian dynamics is essentially a classically dynamics. Same a
conclusion could also be reached from classical physics. With
quantum mechanics, the agreement to the Third Law is found and a
stronger conclusion has been reached: Not only the difference in
entropy should be zero, the entropy itself is zero at zero
temperature.

We may conclude that a complete neglecting noise is not viable
choice in general. When noise is small enough, new phenomena would
happen. Phrasing differently, there appears to exist a bottom near
which there is something.

To summarize, in this section we have shown that except the Third
Law, all other Laws of thermodynamics would follow from Darwinian
dynamics. The concern \cite{sekimoto} on which stochastic
integration method, Ito, Stratonovitch, or others, is consistent
with the Second Law is dissolved: Any of them can be made to be
consistent with the Second Law. We also note that based on the
thermodynamical relations, the fundamental relation of
Eq.(\ref{first}), the conservation of energy of
Eq.(\ref{d-form1}), the universal heat engine efficiency of
Eq.(\ref{efficiency}), supplemented by the additive of extensive
quantities and the temperature of Eq.(\ref{temperature}), the
Boltzmann-Gibbs distribution is implied. In this sense the
statistical mechanics and the thermodynamics are equivalent.

Thermodynamics deals with the steady state properties. The key
property is determined by the Boltzmann-Gibbs distribution of
Eq.(\ref{bg}) which only depends on the Wright evolutionary
potential function $ \phi $ and the "temperature" $ \theta $. The
rest relations are determined by the various symmetries of the
system. No dynamical information can be inferred from them. In
particular, there is no way to recover the information on two
quantities determine the local time scales, the friction matrix $
R $ and the transverse matrix $ T $, from thermodynamics. In this
sense the time is lost in thermodynamics. With this consideration,
it is evident that thermodynamics contains no direction of time
and hence is consistent with the time-reversal conservative
Newtonian dynamics.

\section{ Dynamical Equalities }

We have explored the steady state consequences of the Darwinian
dynamics in statistical mechanics and in thermodynamics. In this
section we explore its general dynamical consequences. Two types
of recently found dynamical equalities will be discussed: one
based on the Feynman-Kac formula and other a generalization of the
Einstein relation.

\subsection{Feynman-Kac formula}

Previous discussions demonstrate that the Boltzmann-Gibbs
distribution plays a dominant role. It is naturally to work in a
representation in which Boltzmann-Gibbs distribution appears in a
most straightforward manner, or, as close as possible. The
standard approach in this spirit is as follows. First, choose the
dominant part of evolution operator $L$. The remaining part is
denoted as $\delta L$. In this subsection a general methodology to
carry out this procedure is summarized.

The Fokker-Planck equation, Eq.(\ref{fp-eq}), can be rewritten as
\begin{equation} \label{fp-eq3}
 \frac{ \partial}{\partial t} \rho({\bf q},t)
  = L (\nabla,{\bf q}; \lambda) \rho({\bf q},t) \, ,
\end{equation}
with $ L = \nabla^{\tau} [D({\bf q}) + A({\bf q})]
   [\theta  \nabla + \nabla \phi({\bf q}) ]$.
It's solution can be expressed in various ways. The most
suggestive form in the present context is that given by Feynman's
path integral \cite{feynman}: If at time $t'$ the system is at
${\bf q}'$, the probability for system at time $t$ and at ${\bf
q}$ is given by summation of all trajectories allowed by
Eq.(\ref{normal}) connection those two points:
\begin{equation}
 \pi({\bf q},t; {\bf q}', t')
  = \left. \left. \sum_{trajectories}
   \right\{ {\bf q}(t) = {\bf q};
            {\bf q}(t') = {\bf q}' \right\} \; .
\end{equation}
In terms of the summation over the trajectories, the solution to
Eq.(\ref{fp-eq3}) (and Eq.(\ref{fp-eq})) may be expressed as
\begin{eqnarray} \label{trajectory}
 \rho({\bf q},t)
  & = & \int d {\bf q}' \, \pi({\bf q},t; {\bf q}', t') \,
   \rho({\bf q},t=0)  \nonumber \\
  & \equiv & \left. \left\langle
     \delta({\bf q}(t)- {\bf q}) \right\rangle \right| _{trajectory} \; .
\end{eqnarray}
The delta function $\delta({\bf q}(t)-{\bf q})$ is used to
explicitly specify the end point. There is a summation over
initial points ${\bf q}'$ weighted by the initial distribution
function $\rho({\bf q}',t=0)$.

Now, considering that the system is perturbed by $\delta L({\bf
q}; \lambda)$, represented, for example, by a change in control
parameter $\lambda$. The new evolution equation is
\begin{equation}
 \frac{ \partial}{\partial t} \rho_{new}({\bf q},t)
  = [ L(\nabla,{\bf q}; \lambda) + \delta L({\bf q};
      \lambda) ] \rho_{new}({\bf q},t) \, .
\end{equation}
The perturbation may act as a source or sink for the probability
distribution. The probability is no longer conserved: in general
$\int d{\bf q} \; \rho_{new}({\bf q},t) \neq \int d{\bf q} \;
\rho_{new}({\bf q},t=0) $. According to the Feynman-Kac formula
\cite{feynmankac}, its solution to this new equation can be
expressed as
\begin{equation} \label{feynmankac}
 \rho_{new}({\bf q},t)
   = \left. \left\langle \delta({\bf q}(t) - {\bf q}) \;
     e^{\int_0^t dt'\; \delta L({\bf q}(t')) }
     \right\rangle \right| _{trajectory} \; ,
\end{equation}
with $ \rho_{new}({\bf q}',t=0) = \rho({\bf q}',t=0)$ and the
trajectories following the dynamics of Eq.(\ref{normal}), the same
as that in Eq.(\ref{trajectory}). Thus, the evolution of the new
density can be expressed by the evolution of the original
dynamics. The corresponding procedure in quantum mechanics is that
in the interaction picture \cite{schiff}. Eq.(\ref{feynmankac}) is
a powerful equality. Various dynamical equalities can be obtained
starting from Eq.(\ref{feynmankac}). Indeed, its direct and
indirect consequences have been extensively explored
\cite{bochkov,evans}.

\subsection{Dynamical work and free energy difference}

We have noticed the special role played by the Botlzmann-Gibbs
distribution, Eq.(\ref{bg}). In particular, it is independent of
the friction and transverse matrices $R,T$. Evidently the
instantaneous Botlzmann-Gibbs distribution with $\lambda =
\lambda(t)$ is
\begin{equation} \label{i-bg}
 \rho_{\theta}({\bf q}; \lambda(t))
  = \frac{ e^{ - \beta \phi({\bf q}; \lambda(t)) } }
         {{\cal Z}_{\theta}(\lambda (0))} \, .
\end{equation}
Here we have explicitly indicated that the parameter is
time-dependent. This distribution function is no longer the
solution of the Fokker-Planck equation of Eq.(\ref{fp-eq}). There
will be transitions out of this instantaneous Boltzmann-Gibbs
distribution function due to the time-dependence of the parameter
$\lambda$. While such transitions may be hard to conceive in
classical mechanics, they can be easily identified in quantum
mechanics, because of discreteness of states \cite{schiff}. One of
such well studied models is the dissipative Landau-Zener
transition \cite{ao1989}.

The interesting question is that whether the transitions can be
reversed such that the instantaneous distribution is indeed an
explicit solution for another but closely related evolution
equation. This means that the original Fokker-Planck equation has
to be modified in a special way to become a new equation. Indeed,
this modified evolution equation can be found for any function
${\bar \rho}({\bf q},t)$, which reads,
\begin{equation} \label{new-eq}
 \frac{ \partial}{\partial t} \rho_{new}({\bf q},t)
  = \left[ L(\nabla,{\bf q},t) - \frac{1}{{\bar \rho}({\bf q},t)}
      ( L(\nabla,{\bf q},t) {\bar \rho}({\bf q},t) )
     + \left( \frac{\partial  \ln | {\bar \rho}({\bf q}, t)  | }
         {\partial t} \right) \right]
     \rho_{new}({\bf q},t) \, .
\end{equation}
It can be verified  $ \rho_{new}({\bf q}, t) = {\bar \rho}({\bf
q},t)$ is indeed a solution of above equation. Treating
$$
 \delta L =
  - \frac{1}{{\bar \rho}({\bf q},t)}
    L(\nabla,{\bf q},t) {\bar \rho}({\bf q},t)
  + \frac{\partial \ln|{\bar \rho}({\bf q}, t)| }
         {\partial t}
$$
and the Feynman-Kac formula Eq.(\ref{feynmankac}) may be applied.
The analogous procedure is well studied on the transitions during
adiabatic processes in interaction picture of quantum mechanics
\cite{schiff,ao1989} and of statistical mechanics \cite{feynman}.

Now, let $\bar \rho$ be the instantaneous Boltzmann-Gibbs
distribution of Eq.(\ref{i-bg}): ${\bar \rho} = \rho_{\theta}({\bf
q}; \lambda(t))$. We have
$$
  \delta L = -\beta  {\dot \lambda} \frac{\partial \phi{\bf q}; \lambda )}
                 {\partial \lambda} \; .
$$
Eq(\ref{new-eq}) can be solved by summing over all trajectories
using the Feynman-Kac formula, Eq.(\ref{feynmankac}). At the same
time, we know the instantaneous Boltzmann-Gibbs distribution of
Eq.(\ref{i-bg}) is its solution. Hence equal those two solutions
to the same equation, we have following equality
\begin{equation} \label{feynman-kac2}
 \frac{ e^{ - \beta \phi({\bf q}; \lambda(t)) } }
      {\int d {\bf q} \, e^{ - \beta \phi({\bf q}; \lambda(0)) }}
  =  \left.  \left\langle
     \delta({\bf q} - {\bf q}(t) ) \;
    e^{ - \beta \int_{0}^{t} d t' \; {\dot \lambda}(t')
     \frac{ \partial \phi({\bf q}(t'); \lambda(t')) }
          {\partial \lambda} } \right\rangle  \right|_{trajectory}
\end{equation}
Following Jarzynski \cite{jarzynski} we define the dynamical work
\begin{equation}
 W_t = \int _{0}^{t} d t' \; {\dot \lambda}(t')
     \frac{ \partial \phi({\bf q}(t'); \lambda(t')) }
          {\partial \lambda}
\end{equation}
The equality between the free energy difference $\Delta F_{\theta}
= F_{\theta}(t) - F_{\theta}(0))$ and the dynamical work $W_t$ is,
after summation over all final points of the trajectories in
Eq.(\ref{feynman-kac2}),
\begin{equation} \label{jarzynski}
 e^{ -\beta \Delta F_{\theta} } =  \left. \left\langle
    e^{ - \beta W_t } \right\rangle  \right|_{trajactory}
\end{equation}
This elegant equality connects the steady state quantities $\Delta
F_{\theta}$ to the work done in a dynamical process. It was first
discovered by Jarzynski \cite{jarzynski}. It should be emphasized
that there is no assumption of steady state state at time $t$ for
the system governed by Eq.(\ref{fp-eq}). In fact, it is known, for
example, in case of the Landau-Zener transition that it is not
\cite{ao1989}. This equality has been discussed and extended by
various authors from various perspectives
\cite{crooks,szabo,sasa,schulten,chernyak,qian}. The connection of
this equality to the Fyenman-Kac formula was first explicitly
pointed out in Ref.\cite{szabo}. There have been experimental
verifications of this equality \cite{collin}.

The Jarzynski equality places the Boltzmann-Gibbs distribution
hence the canonical ensemble in a central position. They are
simply natural consequences from the Darwinian dynamics. However,
if starting from the conservative Newtonian dynamics, the
appropriate ensemble is the micro-canonical ensemble. Any
distribution function which is a function of the potential
function or Hamiltonian would be the solution of the Liouville
equation. From this point of view the Boltzmann-Gibbs distribution
and the associated temperature appear arbitrary: It is just one
among infinite possibilities. This concern has been raised in
literature \cite{cohen} regarding to the generality of the
equality of Eq.(\ref{jarzynski}). No satisfactory treatment of
this concern within Newtonian dynamics has been given. Rather, it
has been an "experimental attitude": If one does this and makes
sure the procedure is correct one gets that, and it works.
Instead, the Darwinian dynamics provides one {\it a priori} reason
to justify the use of Boltzmann-Gibbs distribution in the
derivation of the Jarzynski equality.

\subsection{Generalized Einstein relation}

In deriving the Boltzmann-Gibbs distribution from the Darwinian
dynamics, a generalization of the Einstein relation,
Eq.(\ref{einstein}):
$$
  [R({\bf q}) + T({\bf q}) ] \, D({\bf q}) \,
     [ R({\bf q}) - T({\bf q}) ] = R({\bf q})  \; ,
$$
has been used \cite{ao2004}. This is another general and simple
dynamical equality. In the presence of detailed balance condition,
that is, $T = 0$, this relation reduces to $ R D = 1$, which was
discovered a century ago by Einstein \cite{einstein} and since
known as the Einstein relation. Variants of the Einstein relation
in different settings were obtained earlier and independently by
Nernst \cite{nernst}, Townsend \cite{townsend}, Sutherland
\cite{sutherland}. Similar to the Jarzynski equality, the
generalized Einstein relation is connected to the Boltzmann-Gibbs
distribution.

Experimentally, all those quantities in Eq.(\ref{einstein}) can be
measured. Hence, this generalized Einstein relation should be
subjected to experimental tests in the absence of detailed
balance, that is, when $ T \neq 0$.

For simplicity, we consider a situation realizable with current
technology: a charged nanoparticle or macromolecule, an electron
or a proton, with charge denoted by $e$, in the presence a strong
uniform magnetic field $B$ and emersed in a viscous liquid with
friction coefficient $\eta$. We restrict our attention to two
dimensional case ($n=2$). The corresponding Darwinian dynamical
equation of Eq.(\ref{normal}) in this case is the Langevin
equation with the Lorentz force for a "massless" charged particle
\cite{ashcroft}:
\begin{equation}
 \eta \dot {\bf q} + \frac{e}{c} B \hat{z} \times \dot {\bf q}
   = - \nabla \phi({\bf q}) + N_{II} {\bf \xi}(t)
\end{equation}
The friction matrix is
\begin{equation}
 S = \eta \left( \begin{array}{ll}
           1 & 0 \\
           0 & 1 \end{array}
      \right)
\end{equation}
The transverse matrix is
\begin{equation}
 T = \frac{e}{c} B  \left( \begin{array}{ll}
           0 & 1 \\
           -1 & 0 \end{array}
      \right)
\end{equation}
and the "temperature" is $\theta = k_B T_{BG}$, the Boltzmann
constant and the thermal equilibrium temperature. The
corresponding Fokker-Planck equation following Eq.(\ref{fp-eq}) is
\begin{equation}
 {\partial \rho({\bf q},t) \over \partial t}
  = \nabla [ D \theta \; \nabla
   + [D + A ] \nabla \phi({\bf q})] \rho({\bf q},t) \; .
\end{equation}
This is a precisely a diffusion equation with diffusion matrix
$D$. Both $D$ and $A$ can be obtained via the generalized Einstein
relation, Eq.(\ref{einstein}):
\begin{equation}
 D = \frac{\eta}{\eta^2 + \left( \frac{e}{c} B \right)^2 }
      \left( \begin{array}{ll}
           1 & 0 \\
           0 & 1 \end{array}
      \right)
\end{equation}
\begin{equation}
 A = \frac{ \frac{e}{c} B }{\eta^2 + \left( \frac{e}{c} B \right)^2 }
      \left( \begin{array}{ll}
           0 & -1 \\
           1 & 0 \end{array}
      \right)
\end{equation}
In a typical classical situation, though all quantities can be
measured experimentally, the friction coefficient is likely less
sensitive to the magnetic field. Then the experimentally one may
need to focus on the diffusion in the present of magnetic field
without any potential field. In this case the evolution of
distribution is governed by the standard diffusion equation:
\begin{equation} \label{diffusion}
 {\partial \rho({\bf q},t) \over \partial t}
  =    \theta d_B \; \nabla^2 \rho({\bf q},t) \; ,
\end{equation}
with
$$
 d_B = \frac{ \eta }{ \left[\eta^2 + \left( \frac{e}{c} B \right)^2
       \right] } \, .
$$ The solution to Eq.(\ref{diffusion}) with $\rho({\bf
q},t=0 = \delta {\bf q}(t=0) - {\bf q}$ is standard (two
dimension, $n=2$):
$$
   \rho({\bf q},t) = \frac{ 1 }{2\pi \; t }
    \exp \left\{ - \frac{ {\bf q }^2 }{2d_B \theta \; t } \right\}
$$
Averaging over trajectories governed by Eq.(\ref{diffusion}), $
\left. \langle {\bf q}(t) - {\bf q}(t=0) \rangle
\right|_{trajectory} = 0 $ and
$$
%\begin{equation}
 \left.  \left\langle ({\bf q}(t) - {\bf q}(t=0) )^2 \right\rangle
  \right|_{trajectory} =  4 d_B \theta \; t  \; .
%\end{equation}
$$
The readily experimental system may be that by injection of
electrons into a semiconductor one measures their diffusion in the
presence of a magnetic field. Every quantity in the generalized
Einstein relation of Eq.(\ref{einstein}) can be measured and
controlled experimentally. Such experiments may has already been
done (????). Another experimental system may be on ionized
hydrogen or deuterium. For charged macromolecules and
nano-particles, the friction coefficient may be too large to allow
a measurable magnetic field effect accessible by current magnets.
As a numerical example, for the zero magnetic field diffusion
constant of $d_{B=0} k_B T_{BG} \sim 10^4 \; cm^2/sec$, which
amounts to diffuse about $100 cm$ in 1 second, the friction
coefficient is $\eta = 1/d_{B=0} \sim 4 \times 10^{-16}
dyne/(cm/sec) $ at temperature $T_{BG} = 300 K$.  Assuming one net
electron charge, for magnetic field $ B= 1 \; Telsa$ , we have $eB
/ c \sim 1.6 \times 10^{-16} dyne/(cm/sec) $, comparable to the
friction coefficient.
% They are experimentally accessible(????).

\section{Prospect}

In the present paper we have presented the statistical mechanics
and thermodynamics as natural consequences of the Darwinian
dynamics. Two types of recently found general dynamical equalities
have been explored. Both can be directly tested experimentally.
Everything appears in its right place except one: From the physics
point of view it is the conservative dynamics from which we should
start, not that of the Darwinian. This physics view has indeed
tremendous of experimental supports. Remarkable progresses have
been made along this line of reasoning during past 150 years. It
is still the subject of current intensive research focus
\cite{cohen,lebowitz,gbg}. The physics effort may be condensed to
one question. The natural consequence of the conservative dynamics
is the micro-canonical ensemble, from which the canonical ensemble
just appears to be one of its infinite possibilities. How and why
does Nature choose the canonical ensemble and the Second Law?
There is no consensus yet on the answer .

The difficulty in reaching the Second Law from the conservative
dynamics may give a boost to consider the Darwinian dynamics.
There is, however, a genuine and compelling reason to to do so:
the Darwinian dynamics is the most fundamental and successful
dynamical theory in biological sciences. Furthermore, as having
demonstrated above, from it the Second Law and other
nonequilibrium properties follow naturally. Logically it provides
a simple starting point. It must contain an element of truth.

The conservative dynamics and the Darwinian dynamics appear to
occupy the two opposite ends of the theoretical description of
Nature. Both have been extremely successful. In many aspects they
appear to be complementary to each other. Wether or not there is a
hidden reason such that they are truly related to each other is
not known presently. It waits to be discovered by further
experimental and theoretical studies. The present deliberation may
provide a certain utility for this endeavor.

%
% ensemble vs assemble
% realization of thermodynamics
%  Bose-Einstein, Dirac-Fermi,
%  q-statitics
%

{\ }

{\ }

{\bf acknowledgement}.
We thank
%????
M. Dykman, J. Felsenstein, H. Qian, D.J. Thouless, J. Wang, L.
Yin, X.M. Zhu
%????
for constructive discussions at various stages of this work. There
is a vast body of work done on statistical mechanics and
thermodynamics. No single paper can do justice to the relevant
literature. Admittedly very incomplete, it is my hope a useful
fraction of literature has been covered and a spirit of current
research activities has been captured. In addition, this work is a
critical discussion of two fundamental fields based on an emerging
dynamical formulation. Biases and prejudices are unavoidable. I
apologize to those whose important works are not mentioned here,
likely the result of my own oversight. I would appreciate the
reader's effort very much to bring her/his or other's important
works to my attention (e-mail: aoping@u.washington.edu). This work
was supported in part by USA NIH grant under HG002894.

\end{document}